\documentstyle[twoside,fleqn,espcrc2,epsf]{article}

\title{Review on Double Beta Decay Experiments \& Comparison
       with Theory \protect\thanks{Invited Review talk given
       at the {\em XVIII Intl. Conference on Neutrino Physics and
       Astrophysics}, Neutrino 98, Takayama (Japan), June 1998. To
       be published in Nucl. Phys. B (Proc. Suppl.). Ed. by Y.
       Suzuki, Y. Totsuka.}}

\author{Angel Morales\address{Laboratory of Nuclear Physics and
        High Energy Physics. Faculty of Science, University of Zaragoza, \\
        Pedro Cerbuna 12, 50009 Zaragoza, Spain}}

\begin{document}

\begin{abstract}
\end{abstract}

\maketitle

\section{INTRODUCTION \& MOTIVATION}
In the Standard Model of Particle Physics neutrinos are strictly
massless, although there is no theoretical reason for such prejudice.
On the experimental side, there is not yet conclusive evidence that
the neutrino has a non-zero mass, although the results of several
experiments (widely reported to this Conference) with solar,
atmospheric and terrestrial neutrinos lead to inconsistencies in the
standard theory, unless it is assumed that neutrinos have indeed masses.
Moreover, galaxy formation requires hot (as well as cold) non-baryonic
dark matter to match properly the observed spectral power at all scales
of the universe. A light neutrino of a few eV could make the hot dark
matter, and help to solve the neutrino oscillation problem.

In the Standard Model, neutrinos and antineutrinos are
supposed to be different particles, but no experimental proof has been
provided so far. The nuclear double beta decay addresses both questions:
whether the neutrino is self-conjugated and whether it has a Majorana mass.
In fact, the lepton number violating neutrinoless double beta decay
(A, Z) $\rightarrow$ (A, Z+2) + $2e^{-}$ is the most direct
way to determine if neutrinos are Majorana particles. For this
non-standard $2\beta 0\nu$ process to happen, the emitted neutrino
in the first neutron decay must be equal to its antineutrino and
match the helicity of the neutrino absorbed by the second neutron.
Phenomenologically that implies the presence of a mass term or a
right-handed coupling. A well-known argument of Schechter and Valle
\cite{Sch1} shows that in the context of any gauge theory, whatever mechanism
be responsible for the neutrinoless decay, a Majorana neutrino mass
is required. Moreover \cite{Kay1}, the observation of a $2\beta 0\nu$
decay implies a lower bound for the neutrino mass, i.e. at least one neutrino
eigenstate has a non-zero mass.

Another form of neutrinoless decay,
(A, Z) $\rightarrow$ (A, Z+2) + $2e^{-}+\chi$ may reveal also the
existence of the Majoron ($\chi$), the Goldstone boson emerging
from the spontaneous symmetry breaking of B--L, of most relevance
in the generation of Majorana neutrino masses and of far-reaching
implications in Astrophysics and Cosmology.

These and other issues, like the verification of SUSY models,
compositeness, leptoquarks, etc. make the search for the neutrinoless
double beta decay an invaluable exploration of non-standard model
physics, probing mass scales well above those reached with
accelerators. In this overview we will refer basically
to the question of the neutrino mass in connection with the
current results of the double beta decay searches.

The two-neutrino decay mode (A, Z) $\rightarrow$ \\
(A, Z+2) + $2e^{-}+2\overline{\nu}_{e}$ is a conventional \cite{Hax1},
although rare, second order weak process $(2\beta 2\nu)$,
allowed within the Standard Model. The half-lives are customary expressed as
$[T_{1/2}^{2\nu}\ (0^+\rightarrow 0^+ )^{-1}=G_{2\nu} \mid M_{GT}^{2\nu} \mid^{2}$,
where $G_{2\nu}$ is an integrated kinematical factor \cite{Doi1} and $M_{GT}^{2\nu}$
the nuclear double Gamow Teller matrix element.

The neutrinoless decay half-life (as far as the mass term contribution is
concerned) is expressed as
$(T_{1/2}^{0\nu})^{-1}=F_N<m_\nu>^2/m_e^2$, where
$F_N\equiv G_{0\nu} \mid M^{0\nu} \mid ^2$ is a nuclear factor-of-merit
and $M^{0\nu}$ is the neutrinoless nuclear matrix-element,
$M^{0\nu}=M_{GT}^{0\nu}-(g_V/g_A)^2\ M_F^{0\nu}$, with $M_{GT,F}^{0\nu}$
the corresponding Gamow-Teller and Fermi contributions. $G_{0\nu}$
is an integrated kinematic factor \cite{Doi1}. The quantity
$<m_\nu>=\Sigma \lambda_jm_jU_{ej}^2$ is the so-called effective neutrino
mass parameter, where $U_{ej}$ is a unitary matrix describing the mixing
of neutrino mass eigenstates to electron neutrinos, $\lambda_j$ a CP
phase factor, and $m_j$ the neutrino mass eigenvalue.

As far as the neutrinoless decay with the emission of Goldstone
bosons is concerned, various Majoron models have been invented
to circumvent the $Z^0$ width constrain on the number of neutrino
species---which ruled out the original Majoron models---and
to allow, at an observable rate, double beta neutrinoless decays
with Majoron (or other massless or light bosons) emission
$(2\beta0\nu \chi)$. The $(0\nu \chi)$ half-life is expressed as
$T_{0\nu \chi}^{-1}=<g> \mid M^{0\nu} \mid^2 G_{0\nu \chi}$, where
$M^{0\nu}$ is the same matrix element as in the $2\beta 0\nu$ and
$g$ the Majoron coupling to neutrinos $(g\chi \overline{\nu}_e \gamma_5 \nu_e)$.

Concerning the neutrino mass question, the discovery of a
$2\beta 0\nu$ decay will tell that the Majorana neutrino has a mass
equal or larger than $<m_\nu>=m_e/(F_NT_{1/2}^{0\nu})^{1/2}$ eV,
where $T_{1/2}^{0\nu}$ is the neutrinoless half life. On the contrary,
when only a lower limit of the half-life is obtained (as it is the
case up to now), one gets only an upper bound on $<m_\nu>$, but not
an upper bound on the masses of any neutrino. In fact, $<m_\nu>_{exp}$
can be much smaller than the actual neutrino masses. The $<m_\nu>$
bounds depend on the nuclear model used to compute the $2\beta 0\nu$
matrix element. The $2\beta 2\nu$ decay half-lives measured
till now constitute bench-tests to verify the reliability of the
nuclear matrix element calculations which, obviously, are of paramount
importance to derive the Majorana neutrino mass upper limit.

\section{STRATEGIES FOR DOUBLE BETA DECAY SEARCHES}

The experimental signatures of the nuclear double beta decays are in
principle very clear: In the case of the neutrinoless decay, one should
expect a peak (at the Q$_{2\beta}$ value) in the two-electron summed energy
spectrum, whereas two continuous spectra (each one of well-defined shape)
will feature the two-neutrino and the Majoron-neutrinoless decay modes
(the first having a maximum at about one third of the Q value, and the latter
shifted towards higher energies). In spite of such characteristic imprints,
the rarity of the processes under consideration make very difficult their
identification. In fact, double beta decays are very rare phenomena,
with two-neutrino half-lives as large as $10^{19}$ y to $10^{24}$ y and
with neutrinoless half-lives as long as $10^{25}$ y (and above), as the best
lower limit stands by now. Such remotely probable signals have to be
disentangled from a (much bigger) background due to natural radioactive decay
chains, cosmogenic-induced activity, and man-made radioactivity, which
deposit energy on the same region where the $2\beta$ decays do it but
at a faster rate. Consequently, the main task in $2\beta$-decay searches
is to diminish the background as much as possible by going underground
and using state-of-the-art ultralow background techniques to supress
it or to identify it and subtract it. All the experiments follow this
general strategy because the experimental sensitivity in $2\beta$ decay
searches is limited by the level of background achieved.

To measure $2\beta$ decays, three general approaches have been followed:
The geochemical experiments, where isotopic anomalies in noble gases daughter
of $2\beta$ decaying nucleus over geological time scales are looked for.
Some examples are the decays of $^{82}$Se, $^{96}$Zr, $^{128}$Te, $^{130}$Te.
Another method is that of the radiochemical experiments, which are based on the
fact that when the daughter nuclei of a double beta emitter are themselves
radioactive, they can be accumulated, extracted and counted.
Examples are $^{238}$U, $^{244}$Pu.

Most of the recent activity, however, refers to direct counting experiments,
which measure the energy of the $2\beta$ emitted electrons and so the spectral
shapes of the $2\nu$, $0\nu$, and $0\nu \chi$ modes of double beta decay.
Some experimental devices track also the electrons (and other charged particles),
measuring the energy, angular distribution, and topology of events. The tracking
capabilities are essential to discriminate the $2\beta$ signal from the background.
The types of detectors currently used are:

\begin{itemize}
\item Calorimeters where the detector is also the $2\beta$ source (Ge diodes,
scintillators---CaF$_2$, CdWO$_4$---,thermal detectors, ionization chambers).
They are calorimeters which measure the two-electron sum energy and
discriminate partially signal from background by pulse shape analysis (PSD).
\item Tracking detectors of source$\neq$detector type (Time Projection Chambers
TPC, drift chambers, electronic detectors). In this case, the $2\beta$
source plane(s) is placed within the detector tracking volume, defining
two---or more---detector sectors.
\item Tracking calorimeters: They are tracking devices where the tracking
volume is also the $2\beta$ source. Only one of this type of device is
operating (a Xenon TPC), but there are others in project.
\end{itemize}

Well-known examples of $2\beta$ emitters measured in direct counting
experiments are $^{48}$Ca, $^{76}$Ge, $^{96}$Zr, $^{82}$Se, $^{100}$Mo,
$^{116}$Cd, $^{130}$Te, $^{136}$Xe, $^{150}$Nd.

The strategies followed in the $2\beta$ searches are varied. Calorimeters
of good energy resolution and almost 100\% efficiency (Ge-detectors, Bolometers)
are well suited for $0\nu$ searches. However, they lack the tracking
capabilities to identify the background on an event-by-event basis.
Pulse Shape Discrimination (PSD) will help. Simultaneous measure of
heat and ionization would do it.
The Monte Carlo (MC) modeling of the background spectrum to be subtracted
from the data is approximate. So, one should first reduce the
radioimpurities as much as possible and then trace back and MC-model
the remaining contaminations and subtract them. On the contrary, the identification
capabilities of the various types of chambers make them very well
suited for $2\nu$ and $0\nu \chi$ searches. However, their energy
resolution is rather modest and the efficiency is only of a few percent.
Furthermore, the ultimate major background source in these devices when
looking for $2\beta 0\nu$ decay will be that due to the standard
$2\beta 2\nu$ decay. The rejection of background provided by the tracking
compensates, however, the figure of merit in $0\nu$ searches. Modular calorimeters
can have large amounts of $2\beta$ emitters (Heidelberg/Moscow, IGEX,
CUORE and GENIUS project). However, current operating chambers---except
the Xe/TPC---cannot accommodate large amounts of $2\beta$ emitters in
the source plate. Future tracking devices will have 10 kg and more
(NEMO3, MUNU).

As a general rule, the detector must optimize the so-called detector
factor-of-merit or neutrinoless sensitivity (introduced by the
pioneer work of E. Fiorini), which for source$=$detector devices reads
$F_D=4.17 \times 10^{26}(f/A)(Mt/B\Gamma)^{1/2} \varepsilon_{\Gamma}$
years where B is the background rate (c/keV kg y), M the mass of
$2\beta$ emitter (kg), $\varepsilon_{\Gamma}$ the detector efficiency
in the energy bin $\Gamma$ around Q$_{2\beta}$ ($\Gamma =$ FWHM) and
t the time measurement in years (f is the isotopic abundance and A
the mass number). The other guideline of the experimental strategy is
to choose a $2\beta$ emitter of large nuclear factor of merit
$F_N=G_{0\nu} \mid M^{0\nu} \mid^2$, where the kinematical factor
qualifies the goodness of the $\rm Q_{2\beta}$ value and $M^{0\nu}$
the likeliness of the transition. Notice that the upper limit on
$<m_\nu>$ is given by $<m_\nu><m_e/(F_D F_N)^{1/2}$.

\section{OVERVIEW OF EXPERIMENTAL SEARCHES}

In the following we will overview some of the direct counting experiments,
reporting only on $2\beta$ transitions to the ground
state. A considerable activity has been done recently on transitions
to excited states but we will omit them for lack of space.

There exist two experiments in operation looking for the double beta
decay of $^{76}$Ge. They both employ large amounts of enriched $^{76}$Ge
in sets of detectors. The Heidelberg/Moscow Collaboration experiment
(a set of five large Ge detectors amounting to 10.2 kg) running in
Gran Sasso \cite{Kla1} (exposed by H.V. Klapdor-Kleingrothaus in these Proceedings),
and the IGEX Collaboration in Canfranc (Spain), which is described below.

\begin{table*}[t]
\caption{Theoretical half-lives $T_{1/2}^{2\nu}$ in some
representative nuclear models versus direct experiments.}
\label{tab:tab1}
\begin{tabular*}{\textwidth}{l@{\extracolsep{\fill}}lllllllll}
\hline
                 & \multicolumn{8}{c|}{\sl{Theory}}
                 & \multicolumn{1}{c}{\sl{Experiment}} \\ \hline

                 & \multicolumn{3}{|c|}{SM}
                 & \multicolumn{2}{c|}{QRPA}
                 & \multicolumn{1}{c|}{$1^+D$}
                 & \multicolumn{1}{c|}{OEM}
                 & \multicolumn{1}{c|}{MCM}
                 & \\ \hline

                 & \cite{Hax1}
                 & \cite{Hax2}
                 & \cite{Cau2}
                 & \cite{Moe1}
                 & \cite{Sta1}
                 & \cite{Aba1}
                 & \cite{Wu1}
                 & \cite{Suh2}
                 & \\ \hline

$^{48}$Ca($10^{19}$y) & 2.9 & 7.2 & 3.7 &  &  &  &  &   &
\begin{math}
4.3
\begin{array}{l}
+2.4\\
-1.1
\end{array}
\pm 1.4
\end{math}
UCI \\ \hline

$^{76}$Ge($10^{21}$y) & 0.42 & 1.16 & 2.2 & 1.3 & 3.0 &  & 0.28 & 1.9 &
\begin{math}
1.77
\begin{array}{l}
+0.14\\
-0.12
\end{array}
\end{math}
H/M \\
  &  &  &  &  &  &  &  & & $1.45 \pm 0.15$ IGEX \\ \hline

$^{82}$Se($10^{20}$y) & 0.26 & 0.84 & 0.5 & 1.2 & 1.1 & 2.0 & 0.88 & 1.1 &
\begin{math}
1.08
\begin{array}{l}
+0.26\\
-0.06
\end{array}
\end{math}
UCI \\
  &  &  &  &  &  &  &  &  & $0.83 \pm 0.09 \pm 0.06$ NEMO \\ \hline

$^{96}$Zr($10^{19}$y) &  &  &  & 0.85 & 1.1 &  &  & .14-.96 &
\begin{math}
2.1
\begin{array}{l}
+0.8\\
-0.4
\end{array}
\pm 0.2
\end{math}
NEMO \\ \hline

$^{100}$Mo($10^{19}$y) &  &  &  & 0.6 & 0.11 & 1.05 & 3.4 & 0.72 &
\begin{math}
1.15
\begin{array}{l}
+0.30\\
-0.20
\end{array}
\end{math}
Osaka \\
    &  &  &  &  &  &  &  & &
\begin{math}
1.16
\begin{array}{l}
+0.34\\
-0.08
\end{array}
\end{math}
UCI \\
    &  &  &  &  &  &  &  & &
    $0.95 \pm 0.04 \pm 0.09$ NEMO \\ \hline

$^{116}$Cd($10^{19}$y) &  &  &  &  & 6.3 & 0.52 &  & 0.76 &
\begin{math}
2.6
\begin{array}{l}
+0.9\\
-0.5
\end{array}
\end{math}
Osaka \\

  &  &  &  &  &  &  &  & &
\begin{math}
2.7
\begin{array}{ll}
+0.5 & +0.9 \\
-0.4 & -0.6
\end{array}
\end{math}
Kiev \\
    &  &  &  &  &  &  &  & & $3.75 \pm 0.35 \pm 0.21$ NEMO \\ \hline

$^{136}$Xe($10^{21}$y) &  &  & 2.0 & 0.85 & 4.6 &  &  &  &
$\geq 0.55$ Gothard \\ \hline

$^{150}$Nd($10^{19}$y) &  &  &  &  & 0.74 &  &  &  &
\begin{math}
1.88
\begin{array}{l}
+0.66\\
-0.39
\end{array}
\pm 0.19
\end{math}
ITEP \\
 &  &  &  &  &  &  &  & &
 \begin{math}
0.675
\begin{array}{l}
+0.037\\
-0.042
\end{array}
\pm 0.068
\end{math}
UCI \\ \hline

\end{tabular*}
\end{table*}

The International Germanium Experiment (IGEX) has three large enriched
(up to 86\%) detectors ($\sim 2$ kg) and three smaller ones ($\sim 1$ kg).
The FWHM energy resolutions of the large detectors at 1333-keV are 2.16, 2.37,
and 2.13 keV, and the energy resolution of the summed data is 4 keV (at
the Q$_{2\beta}$ value of 2038 keV). They feature a unique electroformed
copper technology in the cryostat and use ultralow background materials.
The first stage FET (mounted on a Teflon block a few centimetres apart from the
centre contact of the crystal) is shielded by 2.6 cm of 500 y old lead to
reduce the background. Also the protective cover of the FET and the glass
shell of the feedback resistor were removed for such purpose. Further
stages of amplification are located 70 cm away from the crystal. All the
detectors have preamplifiers modified for pulse shape analysis (PSD) for
background identification.

The Canfranc IGEX setup consists in an innermost shield of 2.5 tons ($\sim 60$ cm cube)
of archaeological lead (2000 yr old)---having a $^{210}$Pb($^{210}$Bi) content of
$<0.01$ Bq/kg---, where the 3 large detectors are fitted into precision-machined
holes to minimize the empty space around the detectors available to radon. Nitrogen
gas evaporated from liquid nitrogen, is forced into the remaining free space to
minimize radon intrusion. Surrounding the archaeological lead block there is
a 20-cm thick layer of low activity lead ($\sim 10$ tons), sealed with
plastic and cadmium sheets. A cosmic muon veto and a neutron shield close
the assembly.

The background recorded in the energy region between 2.0 and 2.5 MeV is about
0.2 c/keV kg y prior to PSD. Background reduction through pulse shape
discrimination is in progress to eliminate multisite events, characteristic
of non-$2\beta$ events. This technique is currently capable of rejecting about
one third of the background events, so the current IGEX background is
$\leq$ 0.07 c/keV kg y. Further preamplifier development and pulse shape
simulations are expected to improve the background rejection efficiency,
pursuing the goal of probing Majorana neutrino
masses corresponding to half-lives of $10^{26}$ years. The current results
of IGEX, both for the $2\beta 2\nu$ and $2\beta 0\nu$ decay modes, are given in
Tables \ref{tab:tab1} and \ref{tab:tab2}. The two-electron summed energy
spectrum around $\rm Q_{2\beta}=2038$ keV region is shown in Figure \ref{fig:fig1}
for an exposure of 92.68 mole years. Data from one of the
large detectors---which went underground in Canfranc more than
three years ago---corresponding to 291 days, were used to set a
value for the $2\nu$-decay mode half-life by simply subtracting
MC-simulated background. Figure \ref{fig:fig2ab}a shows the best fit
to the stripped data corresponding to a half-life
$T^{2\nu}_{1/2}=(1.45 \pm 0.20) \times 10^{21}$ y, whereas
Figure \ref{fig:fig2ab}b shows how the experimental points fit the
double beta Kurie plot.

\begin{table}[hbt]
\caption{Limits on Neutrinoless Decay Modes}
\label{tab:tab2}
\begin{tabular}{llcc}
\hline
\sl{Emitter} & \sl{Experiment} & $\sl T_{1/2}^{0\nu}$ & \sl{C.L.} \\ \hline
$^{48}$Ca & HEP Beijing & $>1.1 \times 10^{22}$ y & 68\% \\
$^{76}$Ge & MPIH/KIAE & $>1.2 \times 10^{25}$ y & 90\% \\
          & IGEX  & $>0.8 \times 10^{25}$ y & 90\% \\
$^{82}$Se & UCI & $>2.7 \times 10^{22}$ y & 68\% \\
          & NEMO 2 & $>9.5 \times 10^{21}$ y & 90\% \\
$^{96}$Zr & NEMO 2 & $>1.3 \times 10^{21}$ y & 90\% \\
$^{100}$Mo & LBL/MHC/ & $>2.2 \times 10^{22}$ y & 68\% \\
           & UNM         &                         &      \\
           & UCI & $>2.6 \times 10^{21}$ y & 90\% \\
           & Osaka & $>2.8 \times 10^{22}$ y & 90\% \\
           & NEMO 2 & $>6.4 \times 10^{21}$ y & 90\% \\
$^{116}$Cd & Kiev & $>3.2 \times 10^{22}$ y & 90\% \\
           & Osaka & $>2.9 \times 10^{21}$ y & 90\% \\
           & NEMO 2 & $>5 \times 10^{21}$ y & 90\% \\
$^{130}$Te & Milano & $>7.7 \times 10^{22}$ y & 90\% \\
$^{136}$Xe & Caltech/UN/ & $>4.4 \times 10^{23}$ y & 90\% \\
           & PSI         &                         &      \\
$^{150}$Nd & UCI & $>1.2 \times 10^{21}$ y & 90\% \\ \hline

\end{tabular}
\end{table}

The Time Projection Chamber TPC of the UC Irvine group is a rectangular
box filled with helium and located underground at 290 m.w.e. (Hoover Dam).
A central $2\beta$ source plane divides the volume into two halves.
A magnetic field of 1200 Gauss is placed perpendicular to the source
plane. Electrons emitted from the source follow helical trajectories
from where the momentum and the angles of the $\beta$-particles are
determined. The $2\beta$ signal is recognized as two electron emitted
from a common point in the source with no other associated activity
during some time before and after the event. The $2\beta$ source is
thin enough (few mg/cm$^2$) to allow $\alpha$-particles to escape and
be detected for tagging the background. The UCI TPC has measured the
two-neutrino double beta decay of $^{82}$Se, $^{100}$Mo, $^{150}$Nd
and $^{48}$Ca (this last case in a collaboration with Caltech and the
Kurchatov Institute), with efficiencies of about $\sim 11$\% and energy
resolution of $\sim 10$\% at the Q value. Figures \ref{fig:fig3abc}.1,
\ref{fig:fig3abc}.2 and \ref{fig:fig3abc}.3 show respectively
\cite{Des1} the UCI $2\beta 2\nu$ decay spectra of $^{100}$Mo,
$^{150}$Nd and $^{48}$Ca, depicting in each case the measured spectra and
their background components as well as the corresponding $2\beta$-decay
best fits. Results are quoted in Tables \ref{tab:tab1} and \ref{tab:tab2}.

\begin{figure}[h]
\vspace{9pt}
\framebox[70mm]{\rule[-21mm]{0mm}{40mm}}
\caption{}
\label{fig:fig1}
\end{figure}

\begin{figure}[h]
\vspace{9pt}
\framebox[70mm]{\rule[-21mm]{0mm}{70mm}}
\caption{}
\label{fig:fig2ab}
\end{figure}

The NEMO 2 apparatus \cite{Piq1} is an electron tracking detector (with open
Geiger cells) filled with helium gas. An external calorimeter (plastic
scintillator) covers the tracking volume and measures the $\beta$ energies
and time of flight. The $2\beta$ source is placed in a central vertical
plane and is divided in two halves, one enriched and another of natural
abundance (of about 150 grams each), to monitor and subtract the background.
To identify a $2\beta$ signal, one should have a 2e-track with a common vertex
($cos\alpha <0.6$) in the source plus two fired plastic scintillators
(E deposition$>$200 keV each). The two-electron events are selected by time of
flight analysis (in the energy range of $2\beta$). NEMO 2 has been operating
for several years at the Modane Underground Laboratory (Frejus Tunnel)
at 4800 m.w.e and has measured the $2\beta 2\nu$ decays of $^{100}$Mo,
$^{116}$Cd, $^{82}$Se and $^{96}$Zr (see Figures \ref{fig:fig4abcd}a,b,c,d)
with an efficiency of about $\varepsilon_{2\nu} \sim 2$\% and an energy
resolution $\Gamma$ (1 MeV)$=18$\% (for results refer to Table \ref{tab:tab1}
and Table \ref{tab:tab2}). A new, bigger detector of the NEMO series, NEMO
3, is ready to start running next year, with 10 kg of $^{100}$Mo.

The ELEGANTS V detector of the University of Osaka (placed successively in
Kamioka and Otho) is
an electron tracking detector which consists of two drift chambers for
$\beta$-trajectories, sixteen modules of plastic scintillators for $\beta$
energies and timing measurement, and twenty modules of NaI for X- and
$\gamma$-rays identification. The $2\beta$ signals should appear as
two tracks in the drift

\begin{figure}[h]
\vspace{9pt}
\framebox[70mm]{\rule[-21mm]{0mm}{95mm}}
\caption{}
\label{fig:fig3abc}
\end{figure}

\noindent
chamber with the vertex in the source plus
two signals from two plastic scintillators segments. Both enriched and
natural sources (of about 100 grams) are employed in the detector for
background monitoring and subtraction. This detector has measured
\cite{Eji1} the $2\beta 2\nu$ decay of $^{116}$Cd, $^{100}$Mo (see Figure
\ref{fig:fig5ab}a,b) with efficiencies
of $\varepsilon_{2\nu} \sim 7$\%$\sim 10$\% and
$\varepsilon_{0\nu} \sim 20$\%, and energy resolution of 150 keV at
1 MeV (the results of ELEGANTS V are quoted in Tables \ref{tab:tab1}
and \ref{tab:tab2}). A new variant of ELEGANTS is searching for the
double beta decay of $^{48}$Ca.

\begin{figure}[h]
\vspace{9pt}
\framebox[70mm]{\rule[-21mm]{0mm}{55mm}}
\caption{}
\label{fig:fig4abcd}
\end{figure}

The Caltech/PSI/Neuchatel Collaboration \cite{Vui1} investigates the double
beta decay of $^{136}$Xe in the Gothard Tunnel (3000 m.w.e.) by using
a time projection chamber where the Xenon is at the same time the
source and the detector medium, i.e. a calorimeter plus a tracking
device. It has a cylindrical drift volume of 180 fiducial litres at
a pressure of 5 atm. The Xenon is enriched up to 62.5\% in $^{136}$Xe,
with a total mass of m$=$3.3 kg. The energy resolution is 6.6\% at
2.48 MeV and the $2\beta 0\nu$ efficiency $\varepsilon_{0\nu} \sim 30$\%.
The $2\beta$ signal appears as a continuous trajectory with distinctive
end features: a large angle multiple scattering and increase charge
deposition (charge ``blobs'') at both ends. As usual, the $2\beta$
topology gives powerful background rejection, leading to a figure
of $\rm B \sim 10^{-1}-10^{-2}$ c/keV kg y (at 2480 keV).
In the neutrinoless decay mode search, the experimental set up has
already reached its limit (Table \ref{tab:tab2} and Figure \ref{fig:fig8_6}).
In the two-neutrino decay mode, the comparison of the single-electron
and two-electron background spectra before and after a recent upgrading
\cite{Vui1} of the readout plane (a factor 4 reduction in single e$^-$
background above 1800 keV) shows that the two-electron spectrum is
not reduced as much as the single-electron one. That implies that
a significant $2\beta$ signal is contained in the 2e data, and so
a new run (at low pressure) is in progress in a search for the
$2\beta 2\nu$ mode.

\begin{figure}[h]
\vspace{9pt}
\framebox[70mm]{\rule[-21mm]{0mm}{83mm}}
\caption{}
\label{fig:fig5ab}
\end{figure}

The ITEP group has measured \cite{Art1}the double beta decay of $^{150}$Nd
(40 g) with a TPC of $\sim 300$ litres filled with CH$_4$ at
atmospheric pressure, in a 700 gauss magnetic field. The detection
efficiency is $\varepsilon_{2\nu} \sim 3$\% (see results in Table
\ref{tab:tab1}). A large ($13 \rm m^3$) TPC is underway for Xe (7.5 kg)
and Nd (5 kg).

\begin{figure}[h]
\vspace{9pt}
\framebox[70mm]{\rule[-21mm]{0mm}{63mm}}
\caption{}
\label{fig:fig8_6}
\end{figure}

The group of INR at Kiev \cite{Dan1}is investigating the double beta decay of
$^{116}$Cd with cadmium tungstate ($^{116}$CdWO$_4$) scintillator
crystals of 12 to 15 cm$^3$ which feature an energy resolution of
$\Gamma =7$\% at 2614 keV. A series of test experiments to reduce
the background has lead to a figure of B$\sim 0.6$ c/keV kg y.
Results are quoted in Tables \ref{tab:tab1} and \ref{tab:tab2}.

A series of bolometer experiments have been carried out by the Milan
group since 1989 in the Gran Sasso Laboratory, searching for the double
beta decay of $^{130}$Te \cite{Cre1}.
The increase of the temperature produced by the energy
released in the crystal due to a nuclear event (i.e. $2\beta$), is
measured by means of a sensor in thermal contact with the
absorber.
The Milan group uses Tellurium oxide crystals as absorbers, and glued NTD
Ge thermistors as sensors. Notice that natural Tellurium contains 34\% of
$^{130}$Te. After using successively TeO$_2$ crystals of 73 g and 334 g,
as well as a set of four of these large crystals, a tower-like array of
20 crystals of 340 g in a copper frame is currently taking data at a
temperature of $\sim 10$ mK. In a recent run, featuring
an energy resolution (summed over the twenty energy spectra) of $\sim 10$ keV at
2615 keV, and a background of about 0.5 c/keV kg day in that region, they got in
only a few days a better neutrinoless half-life limit than in all their previous
experiments (See Table \ref{tab:tab2}). The calibration spectrum of the
summed twenty-crystal spectra and the background around the $\rm Q_{2\beta}$
region corresponding to a short running have been presented to this Conference
\cite{Cre1} and are shown in Figures \ref{fig:fig9_7}a,b. An enlarged version of
this experiment, CUORE (a Cryogenic Underground Observatory for Rare Events)
consisting of an array of 1000 crystals of TeO$_2$ of 750 g each (with NTD
Ge sensors), operating at $7 \sim 10$ mK, is planned to be installed at
Gran Sasso \cite{Cre1}.

\begin{figure}[h]
\vspace{9pt}
\framebox[70mm]{\rule[-21mm]{0mm}{97mm}}
\caption{}
\label{fig:fig9_7}
\end{figure}

\section{EXPERIMENTAL RESULTS CONFRONT THEORY}

Two main lines have been followed in computing the $2\beta$-decay
nuclear matrix elements: Shell Model (SM) and Quasiparticle Random
Phase Approximation (QRPA). Both approaches have been widely applied
with various degrees of success. The current theoretical predictions
of the $2\nu$ decay modes have provided a general framework of
concordance with the experiment (within a factor 2--5). That gives
confidence in the reasonable reliability of the $2\beta 0\nu$ decay
matrix elements used to extract $<m_{\nu}>$ bounds.

The first attempts to calculate $2\beta$ nuclear matrix element
were made by using the nuclear shell model, but as most $2\beta$
emitters are heavy or medium heavy nuclei, it was necessary to use
a weak coupling limit shell model \cite{Hax1,Hax2} and/or truncation of the
model space to cope with the calculation. Such truncations excluded
configurations relevant for the final results. Predictions of such
former calculations are given in Tables \ref{tab:tab1} and \ref{tab:tab3}.
Until recently, large SM calculations were possible only in the
case of $^{48}$Ca \cite{Cau2}a. New progress in SM codes have
allowed to perform large model space SM calculations \cite{Cau2}b
in heavy and medium heavy
nuclei using realistic single particle basis. Still there are important
truncations because of the large valence space. For $^{48}$Ca, $^{76}$Ge
and $^{82}$Se, the results are in good agreement with the experiment,
whereas for $^{136}$Xe there exists some discrepancy
(See Table \ref{tab:tab1}). Estimates of the neutrinoless decays in
this large model space SM calculation give longer neutrinoless decay
half-lives (for equal $<m_{\nu}>$ values) than the QRPA results.

\begin{table*}[hbt]
\caption{Neutrinoless half-lives in various Theoretical Models
(for the $<m_\nu>$ Term) $T_{1/2}^{0\nu}\ <m_\nu>^2$ values are
given in $10^{24}$ (eV)$^2$ y.}
\label{tab:tab3}
\begin{tabular*}{\textwidth}{l@{\extracolsep{\fill}}lllllllll}
\hline
  & $^{76}$Ge  & $^{82}$Se & $^{100}$Mo & $^{128}$Te & $^{130}$Te
  & $^{136}$Xe & $^{150}$Nd & $^{116}$Cd & $^{48}$Ca \\ \hline
  Weak Coupl. SM \cite{Hax1,Hax2} & 1.67 & 0.58 &    &  4.01 & 0.16 &  &  &  &  \\
  $g_A=1.25 (g_A=1)$ & (3.3) & (1.2) &  & (7.8) & (0.31) &  &  &  &   \\ \hline
  Large Space SM \cite{Cau2} & 17.5 & 2.39 & & & & 12.1 & & & 6.25 \\ \hline
  QRPA \cite{Moe1} & 14 & 5.6 & 1.9 & 15 & 0.66 & 3.3 &  &  &  \\ \hline
  QRPA \cite{Sta1} & 2.3 & 0.6 & 1.3 & 7.8 & 0.49 & 2.2 & 0.034 & 0.49 &  \\ \hline
  QRPA \cite{Tom1} & 2.15 & 0.6 & 0.255 & 12.7 & 0.52 & 1.51 & 0.045 &  &  \\ \hline
  OEM \cite{Wu1} & 2.75 & 0.704 &  & 12.6 & 0.723 & 4.29 & 0.056 & 0.583  & \\ \hline
  QRPA with & 18.4 & 2.8 & 350 & 150 & 2.1 & 2.8 &  & 4.8 & 28 \\
  (without) np pair. \cite{Pan1} & (3.6) & (1.5) & (3.9) & (19.2) & (0.86) &  & (4.7)& (2.4) \\
  \hline
\end{tabular*}
\end{table*}

QRPA is simple from a computing point of view; it includes
many features of the two-body interaction which plays a relevant
role in $2\beta$ decays; it is very sensitive to the $J=1^+$,
$T=0$ particle-particle interaction and have contributed significantly
to understand the large suppression of the experimental rates which
failed to be explained by the earlier theoretical approaches.
QRPA was first applied to compute the $2\beta 2\nu$ matrix elements
by the Caltech group \cite{Moe1} using a zero range force. Results in
agreement with experiment were obtained for various $2\beta$
measured decays, when the value of the strength $\rm g_{\rm pp}$
of the particle-particle interaction used was the one fitting
the $\beta^+$ decay of nuclei with magic number of neutrons.
Subsequent works of the groups of Tubingen \cite{Tom1} and of Heidelberg
\cite{Sta1} (both in $2\nu$ and $0\nu$ decays) confirmed and
refined the results with more realistic NN interactions.
The suppression of the $2\beta 2\nu$ matrix elements is extremely
sensitive to the strength $\rm g_{\rm pp}$ of the particle-particle
interaction, which in fact may lead to almost null matrix elements
for values of $\rm g_{\rm pp}$ in its physical range. The great
sensitivity of $M_{GT}^{2\nu}$  on $\rm g_{\rm pp}$ makes difficult
to make definite rate prediction, contrary to the Shell Model case.
The value of $\rm g_{\rm pp}$ has to be adjusted, otherwise the
QRPA rates span a wide range of values. On the contrary,
the $(2\beta)0\nu$ rates are not so sensitive. The neutrino potential
makes the difference with the $(2\beta)2\nu$ case. The various
multipolarities (besides $J^{\pi}=1^+$) arising because of its radial
dependence, wash out much of the suppression. The nuclear sensitivity
of the $0\nu$ rates is rather smooth and the predictions are much
more reliable. The QRPA has been applied to most of the $2\beta$
emitters.

Several QRPA variants (like the Multiple Commutator Method, MCM
\cite{Suh2}) or extensions have been also applied, as
well as some alternative methods, like the Operator Expansion
Method (OEM) \cite{Wu1}, the SU(4) symmetry, the $1^+$ intermediate state
dominance model ($1^+$D) \cite{Aba1}, the pseudo SU(3), and quite a
few more (see Ref. \cite{Suh2} for
a recent theoretical review). The OEM, for instance, which
avoids summation over intermediate states, predicted results much
less sensitive to $g_{pp}$, but has also several drawbacks. The
alterntative $1^+$D model of Zaragoza/Osaka \cite{Aba1,Eji1}, suggested
a long time ago \cite{Aba1}, relies on the fact that in a double beta transition,
the intermediate state (odd-odd nucleus) having $1^+$ ground state
(gs) can decay by EC to the initial gs, and by $\beta^-$ to the
gs of the final nucleus and so the feeding of pertinent ft-values
provide the $2\beta$ decay nuclear matrix elements.
An archetypical example is provided by the transition
$^{100}$Mo--$^{100}$Tc--$^{100}$Ru, which in most of the calculations
is predicted to decay faster than observed ($\sim 10^{19}$ y).
The QRPA did not work either for $^{100}$Mo, nor did some of
their cures like the OEM (almost insensitive to $\rm g_{\rm pp}$),
which fall a factor three apart from the experimental value.
However, by assuming a dominant contribution of the lowest state
of the intermediate nuclei, the correct value of $M_{2\nu}$
could be reproduced \cite{Gri1}, as already noted quite a few years
ago \cite{Aba1} in this and other transitions. Working out this model
(i.e. feeding the single GT transition matrix element as given
by experiment [say from $\beta^-$ and EC decays and/or (pn),(np),
($^3$He,t) reactions, presently being carried out at RCNP
(Osaka)], Ejiri et al. obtained $2\beta 2\nu$ half-life values
in fair agreement with the experiment \cite{Piq1,Eji2}.

\begin{figure}[h]
\vspace{9pt}
\framebox[70mm]{\rule[-21mm]{0mm}{55mm}}
\caption{}
\label{fig:fig6_8}
\end{figure}

Results of $2\nu$ and $0\nu$ theoretical half-lives are given in Tables
\ref{tab:tab1} and \ref{tab:tab3} according to various nuclear models.
The reader can derive by himself from Tables \ref{tab:tab3} and
\ref{tab:tab2} the $<m_\nu>$ upper bounds according to his preferred
nuclear model. Figure \ref{fig:fig6_8} sumarizes the confrontation
theory vs. experiment in the two-neutrino decay modes, whereas
Figure \ref{fig:fig7_9} gives a comparison of the neutrinoless
half-live results of the various major experiments, and the
corresponding neutrino mass bounds. The white hystograms represent
the half-life limit each experiment must reach to match the
current bounds for the neutrino mass obtained in germanium experiments.

\begin{figure}[t]
\vspace{9pt}
\framebox[70mm]{\rule[-21mm]{0mm}{57mm}}
\caption{}
\label{fig:fig7_9}
\end{figure}

\section{CONCLUSIONS AND OUTLOOK}

The standard $2\beta 2\nu$ decay has been directly observed in several nuclei:
$^{48}$Ca, $^{76}$Ge, $^{82}$Se, $^{96}$Zr, $^{100}$Mo, $^{116}$Cd and
$^{150}$Nd and others are under investigation ($^{130}$Te, $^{136}$Xe).
QRPA reproduces reasonably well the measured half lives with some fine
tuning of the nuclear parameter. Recent (large model space) shell model
calculations give also good predictions in the cases where they have been
applied so far, reinforcing the confidence on the matrix elements needed to
extract the experimental limits on $2\beta 0\nu$ decay. Data from the most
sensitive experiments on $2\beta 0\nu$ lead to the limit
$<m_\nu><0.4-1.5$ eV for the effective neutrino mass, according to
the nuclear model.

The Ge experiments provide the stringest bound to the neutrino mass parameter
and they seem to offer, for the next future, the best prospectives to reach
the lowest values of $<m_\nu>$. The Heidelberg-Moscow experiment and IGEX on
$^{76}$Ge will continue the data taking with a background reduced by pulse
shape discrimination. These experiments will achieve sensitivities of order
$T_{1/2}^{0\nu} \approx 5 \times 10^{25}$ y or close to $10^{26}$ in
$^{76}$Ge, corresponding to $<m_\nu> \approx 0.2-0.6$ eV (according to
the nuclear matrix element used). As proved by the
20-crystal array bolometers of the Milan group, the low temperature thermal
detection of $2\beta$ decays is now mastered. The cryogenic detectors are
supposed to provide better energy resolution and more effective absorption of
the particle energy (thermal vs ionization) and so the CUORE project
is a promising (and feasible) undertaking.

Summarizing, currently running or planned experiments (H/M, IGEX, NEMO 3, MUNU,
Bolometer Arrays), will explore effective neutrino masses down to about 0.1---0.3
eV. To increase the sensitivity it is necessary to go to larger source masses
and reduce proportionally the background. That would bring the sensitivity
to neutrino mass bounds below the tenth of electronvolt. Projects like
CUORE \cite{Cre1} or GENIUS \cite{Kla1} go in that direction.
To pursue the goal even further, huge detector masses, and still better event
identification are needed.

In spite of the progress, a long way is still ahead of us. What is at stake
is worth the effort.

\section{ACKNOWLEDGEMENTS}

I am indebted to my colleagues of the IGEX Collaboration,
in particular to J. Morales for discussion and comments, and to CICYT
(Spain) and the Commission for Cultural, Education and Scientific Exchange
between the United States of America and Spain for financial support.

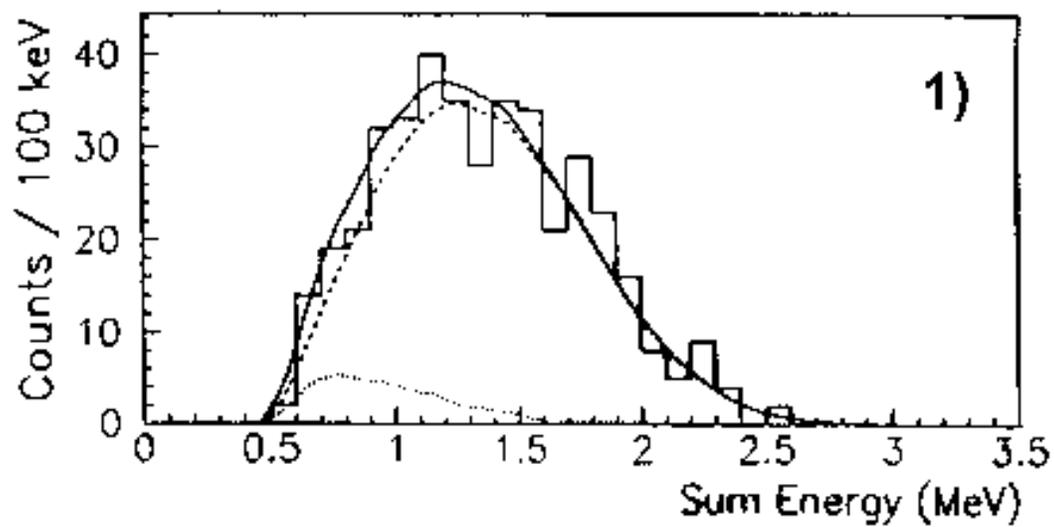
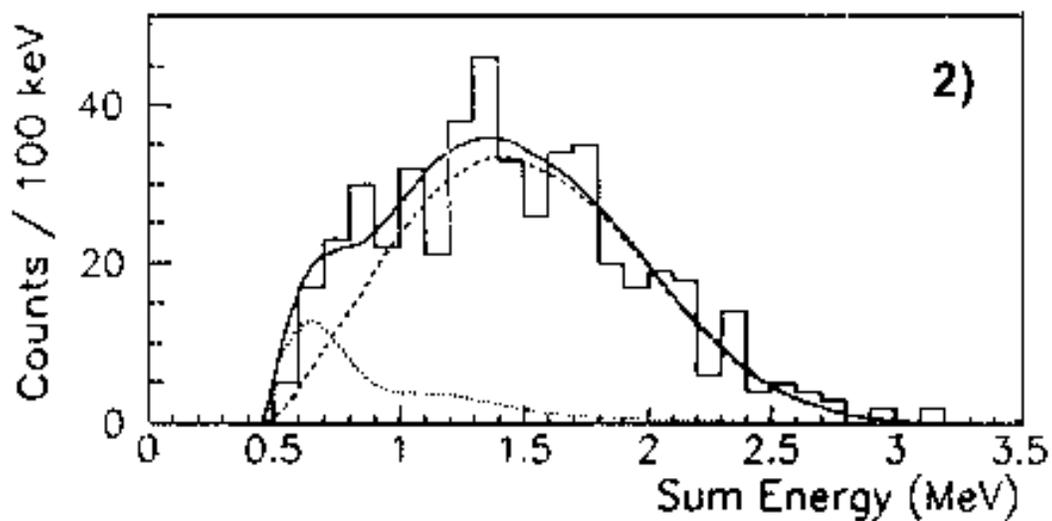
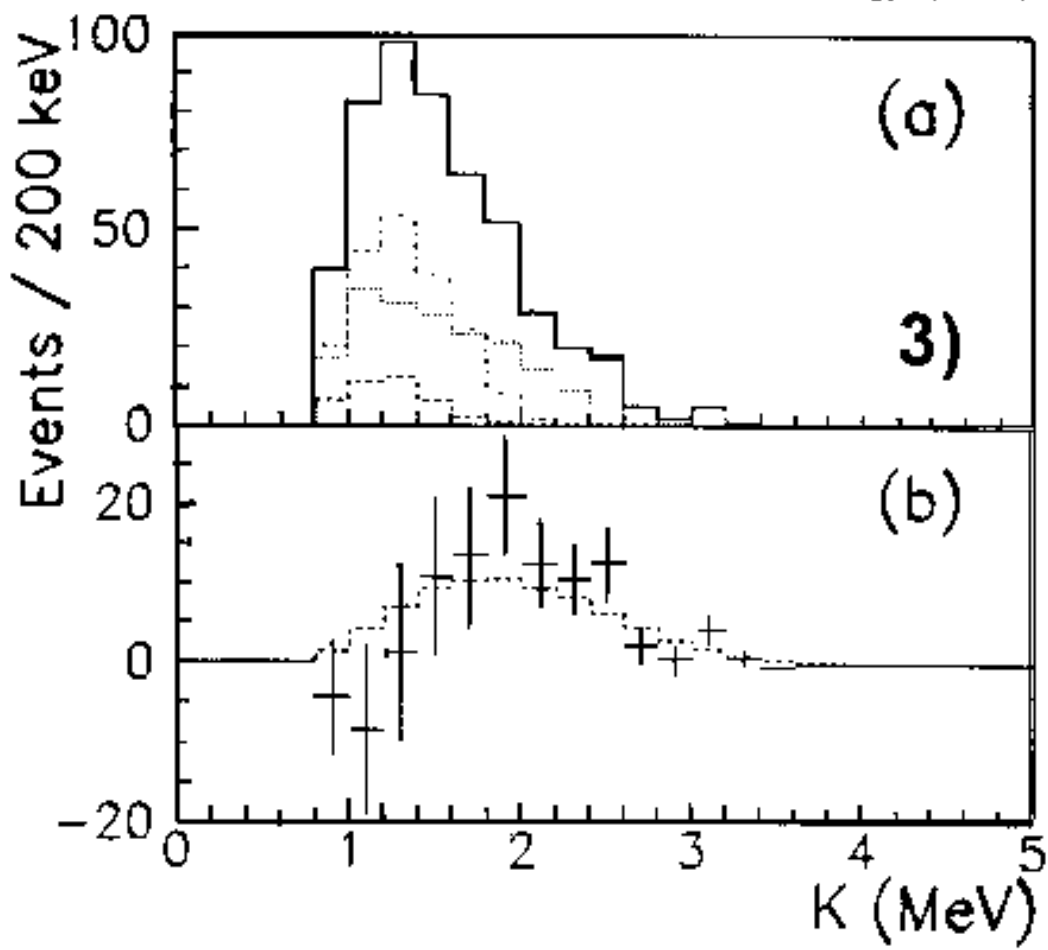
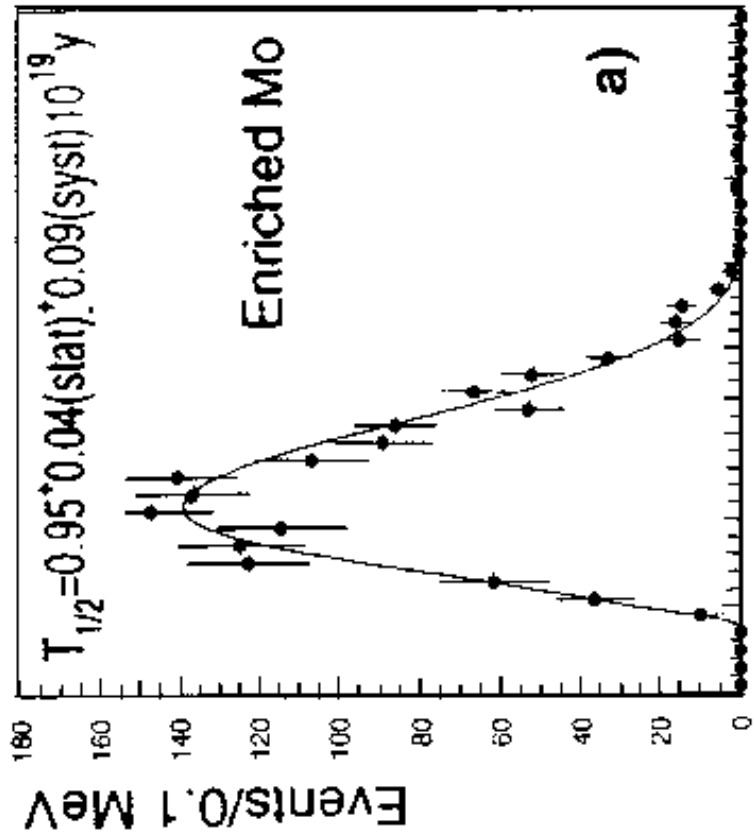
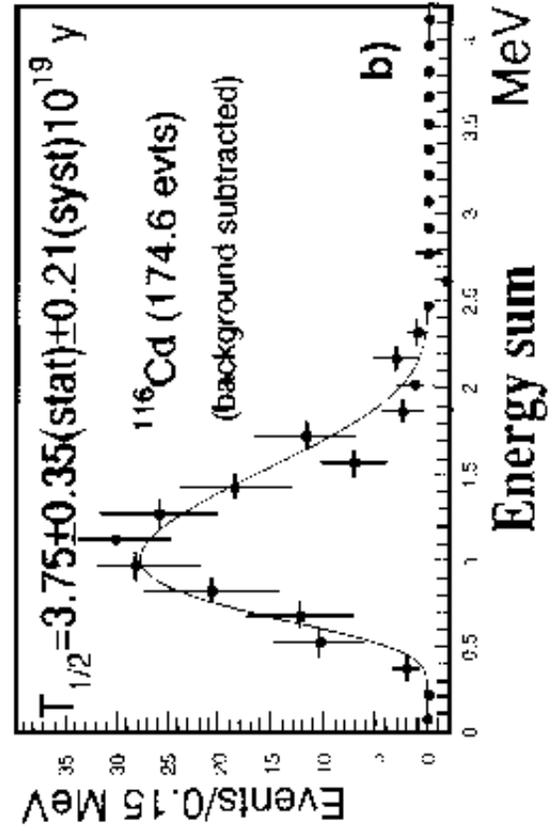
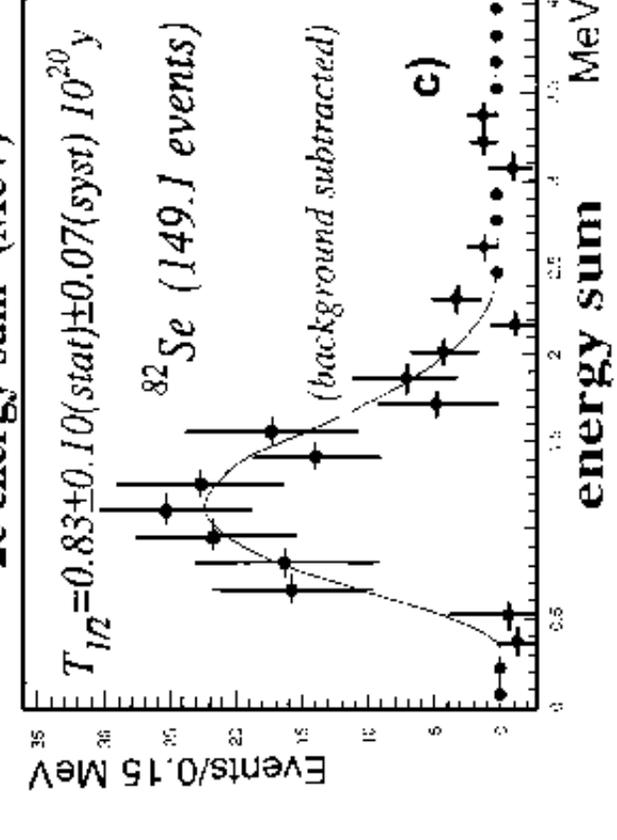
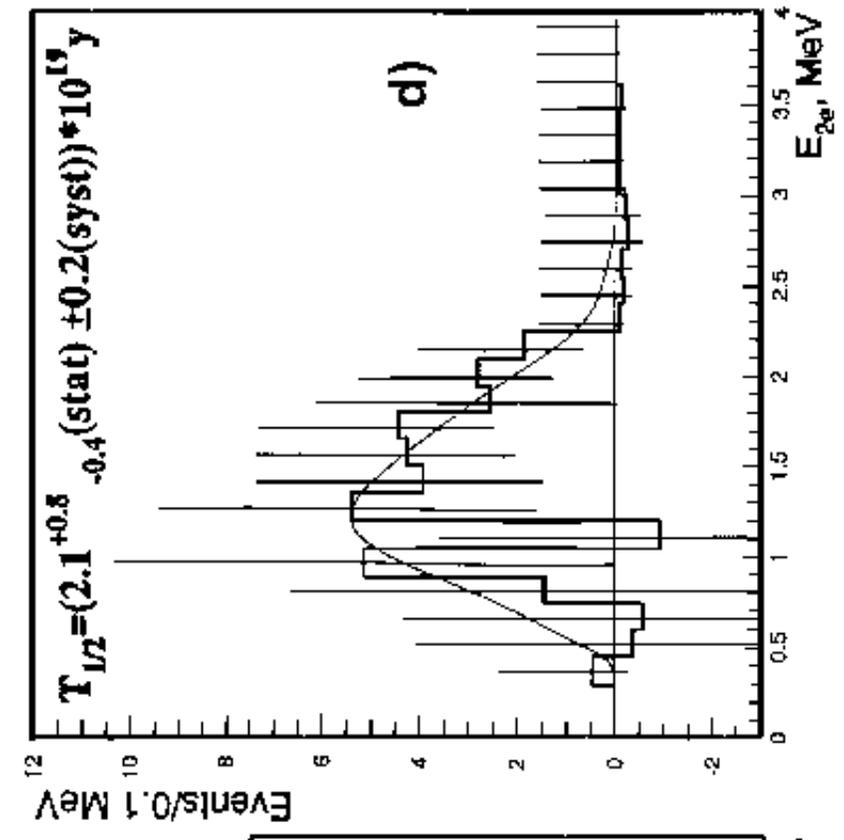
\end{document}